\documentclass{ws-procs9x6}

\begin{document}

\title{Towards a determination of the spectrum of QCD using a 
       space-time lattice}
\author{K.J.~JUGE, A.~LICHTL and C.~MORNINGSTAR\footnote{\uppercase{S}peaker.}}
\address{Department of Physics, Carnegie Mellon University,
         Pittsburgh, PA 15213, USA}
\author{R.G.~EDWARDS and D.G.~RICHARDS}
\address{Thomas Jefferson National Accelerator Facility,
         Newport News, VA 23606, USA}
\author{S.~BASAK and S.~WALLACE}
\address{Department of Physics, University of Maryland,
         College Park, MD 20742, USA}
\author{I.~SATO}
\address{Lawrence Berkeley Laboratory, 1 Cyclotron Road,
         Berkeley, CA 94720, USA}
\author{G.T.~FLEMING}
\address{Sloane Physics Laboratory, Yale University, 
         New Haven, CT 06520, USA}
\maketitle
\abstracts{
Progress by the Lattice Hadron Physics Collaboration in determining the 
baryon and meson resonance spectrum of QCD using Monte Carlo methods with 
space-time lattices is described.  The extraction of excited-state
energies necessitates the evaluation of correlation matrices of sets of 
operators, and the importance of extended three-quark operators to capture 
both the radial and orbital structures of baryons is emphasized.
The use of both quark-field smearing and link-field smearing in the
operators is essential for reducing the couplings of the operators to
the high-frequency modes and for reducing statistical noise in the correlators.
The extraction of nine energy levels in a given symmetry channel
is demonstrated, and identifying the continuum spin quantum numbers of 
the levels is discussed.
}

\section{Introduction}
A charge from the late Nathan Isgur to use Monte Carlo methods to extract the
spectrum of hadron resonances and hadronic properties resulted in the formation
of the Lattice Hadron Physics Collaboration (LHPC) in the year 2000.  As part
of a national collaboration of lattice QCD theorists, the LHPC acquired
funding from several sources, including the DOE's Scientific Discovery
through Advanced Computing initiative, to build large computing clusters at 
JLab, Fermilab, and Brookhaven, as well as to develop the software to carry out
the needed large-scale computations.  The LHPC has several broad goals:
to compute the spectrum of QCD from first principles, to investigate 
hadron structure by computing form factors, structure functions, and other
matrix elements, and to study hadron-hadron interactions.  This talk
focuses solely on our efforts to determine the spectrum of QCD.

Extracting the spectrum of resonances in QCD is a big challenge.
The determination of excited-state energies requires the use of
correlation matrices of sets of operators.  The masses and widths of unstable
hadrons (resonances) must be deduced from the energies of single-particle
and multi-hadron stationary states in a finite-sized box.  This necessitates
the use of multi-hadron operators in the correlation matrices, and the 
computations must be performed in full QCD at realistically-light quark masses.
For these reasons, the computation of the QCD spectrum is indeed a long-term
project.  This talk is a brief status report of our efforts.  

I first describe in Sec.~\ref{sec:how}
how excited-state energies are extracted in our Monte Carlo calculations,
and discuss issues related to unstable resonances.  Our construction
of baryon operators is then outlined in Sec.~\ref{sec:ops} (see
Ref.~\refcite{lhpc} for further details), and in Sec.~\ref{sec:smear}, 
the importance of using smeared quark and gluon fields is demonstrated.
(see also Ref.~\refcite{lat2005}). The first-time extraction of \textit{nine}
nucleon energy levels is also presented in this section.  Concluding remarks
are made and future work is outlined in the concluding Sec.~\ref{sec:conclude}.

\section{Excited states and resonances}
\label{sec:how}
In the path integral formulation of quantum field theory with imaginary
time, stationary-state energies are extracted from the asymptotic decay
rates of temporal correlations of the field operators. If $\Phi(t)$ is
a Heisenberg-picture operator which annihilates the hadron of interest at 
time $t$, then its evolution $\Phi(t)=e^{Ht}\Phi(0)e^{-Ht}$, where $H$ is
the Hamiltonian, can be used to show that in a finite-sized box,
\begin{eqnarray}
 C(t)&\equiv& \langle 0\vert \Phi(t)\Phi^\dagger(0) \vert 0\rangle
=\sum_n \langle 0\vert e^{Ht}\Phi(0)e^{-Ht} \vert n\rangle\langle n\vert
 \Phi^\dagger(0) \vert 0\rangle,\\
&=& \sum_n \vert\langle 0\vert \Phi(0) \vert n\rangle\vert^2
 e^{-(E_n-E_0)t}=\sum_n A_n e^{-(E_n-E_0)t},
\end{eqnarray}
where $\{\vert n\rangle\}$ is the complete set of discrete eigenvectors of 
$H$.  We assume the existence of a transfer matrix, and temporal boundary
conditions have been ignored for illustrative purposes.  One can then
extract $A_1$ and $E_1-E_0$ as $t\rightarrow\infty$, assuming
$\langle 0\vert\Phi(0)\vert 0\rangle=0$ and $\langle 0\vert\Phi(0)\vert 
1\rangle\neq 0$.

A convenient visual tool for demonstrating energy-level extraction is the
so-called ``effective mass'' defined by $m_{\rm eff}(t)=\ln[ C(t)/C(t+a_t)]$,
where $t$ is time and $a_t$ is the temporal lattice spacing.  The effective
mass tends to the actual mass (or energy) of the ground state as $t$
becomes large, signalled by a plateau in the effective mass.  At smaller
times before this plateau is observed, the effective mass varies due to
contributions from other states in the spectrum.  The correlation $C(t)$
is estimated with some statistical uncertainty since the Monte Carlo method 
is used, and usually, the ratio of the noise to the signal increases with 
temporal separation $t$.
Hence, judiciously-chosen operators having reduced couplings with
contaminating higher-lying states are important in order to observe a
plateau in the effective mass before noise swamps the signal.  Key
ingredients in constructing such operators are the use of smeared quark
and gluon fields, the incorporation of spatially-extended assemblages of
the fields, and the use of sets of different operators to exploit 
improvements from variational methods.

Methods of extracting excited-state energies are well 
known\cite{cmichael,luscherwolff}.  For a given $N\times N$
Hermitian matrix of correlations $C_{\alpha\beta}(t) =  \langle 0\vert \Phi_\alpha(t)
 \Phi^\dagger_\beta(0)\vert 0 \rangle,$ the $N$ \textit{principal correlators}
$\lambda_\alpha(t,t_0)$ are defined as the eigenvalues of the matrix
$C(t_0)^{-1/2}C(t)C(t_0)^{-1/2}$, where $t_0$ is some reference time (typically small),
and one can show that 
\begin{equation}
\lim_{t\rightarrow\infty}\lambda_\alpha(t,t_0)
=e^{-(t-t_0)E_\alpha}(1+O(e^{-t\Delta E_\alpha})),\qquad
\Delta E_\alpha=\min_{\beta\neq \alpha}\vert E_\beta-E_\alpha\vert,
\end{equation}
assuming $E_0=0$ and $\lambda_1\geq\lambda_2\geq\lambda_3\cdots$. 
The $N$ principal effective masses 
$m_\alpha^{\rm eff}(t)=\ln[\lambda_\alpha(t,t_0)/
\lambda_\alpha(t+a_t,t_0)]$ now tend (plateau) to the $N$ lowest-lying
stationary-state energies which couple to the $N$ operators.
The associated eigenvectors are orthogonal, and a
knowledge of the eigenvectors can yield information about the partonic
structure of the states.

Many of the hadron states we wish to study are unstable resonances.  Our
computations are done out of necessity in a box of finite volume with
periodic boundary conditions.  Hence, the momenta of the particles we study
are quantized, so all states are discrete in our computations.  Thus, we
can only determine the discrete energy spectrum of stationary states in
a periodic box, which are admixtures of single hadrons and multi-hadron
states.  Resonance masses and widths must somehow be deduced from the
finite-box spectrum\cite{dewitt,wiese88,luscher91B,rum95}.  Once the masses of
the stable single particle states have been determined, the placement
and pattern of their scattering states are known approximately, and the
dependences of their energies on the volume are roughly known.  Resonances
show up as extra states with little volume dependence.  Our initial goal
is simply to ferret out these resonances, not to pin down their properties
to high precision.

\section{Operator construction}
\label{sec:ops}

Our approach to constructing hadron operators is to directly combine the
physical characteristics of hadrons with the symmetries of the lattice
regularization of QCD used in simulations.  For baryons at rest, our 
operators are formed using group-theoretical projections onto the irreducible
representations (irreps) of the $O_h$ symmetry group of a three-dimensional cubic lattice.
There are four two-dimensional irreps $G_{1g}, G_{1u}, G_{2g}$, $G_{2u}$
and two four-dimensional representations $H_g$ and $H_u$. 
The continuum-limit spins $J$ of our states must be deduced by examining
degeneracy patterns across the different $O_h$ irreps 
(see Table~\ref{tab:irreps}).  For example, a $J^P=\frac{1}{2}^+$ state
will show up in the $G_{1g}$ channel without degenerate partners in the
other channels, and a $J^P=\frac{3}{2}^+$ state will show up in the $H_g$
channel without degenerate partners in the other channels.  Four of the
six polarizations of a $J^P=\frac{5}{2}^+$ state show up as a level in the $H_g$
channel, and the other two will occur as a degenerate partner in the $G_{2g}$ 
channel, whereas three degenerate levels, one in each of the three 
$G_{1g}, G_{2g}, H_g$ channels, may indicate a single $J^P=\frac{7}{2}^+$
state (or the accidental degeneracy of a spin-$\frac{1}{2}$ and a $\frac{5}{2}$
state).

\begin{table}[tb]
\tbl{Continuum limit spin identification:
   the number $n_\Lambda^J$ of times that the $\Lambda$ irrep.\ of the
   octahedral point group $O_h$ occurs in the
   (reducible) subduction of the $J$ irrep.\ of $SU(2)$.  The numbers
   for $G_{1u},G_{2u},H_u$ are the same as for $G_{1g},G_{2g},H_g$,
    respectively.}{
\begin{tabular}{cccc@{\hspace{5em}}cccc} \hline
{} &{} &{} &{} &{}\\[-1.5ex]
  $J$  & $n^J_{G_{1g}}$ & $n^J_{G_{2g}}$ & $n^J_{H_g}$ &
 $J$  & $n^J_{G_{1g}}$ & $n^J_{G_{2g}}$ & $n^J_{H_g}$\\[1ex] \hline
 {} &{} &{} &{} &{}\\[-1.5ex]
  $\frac{1}{2}$  &  $1$ & $0$ & $0$ &$\frac{9}{2} $ &  $1$ & $0$ & $2$ \\[1ex]
  $\frac{3}{2}$  &  $0$ & $0$ & $1$ &$\frac{11}{2}$ &  $1$ & $1$ & $2$ \\[1ex]
  $\frac{5}{2} $ &  $0$ & $1$ & $1$ &$\frac{13}{2}$ &  1 & 2 & 2 \\[1ex]
  $\frac{7}{2} $ &  $1$ & $1$ & $1$ &$\frac{15}{2}$ &  1 & 1 & 3 \\[1ex] \hline
\end{tabular}
\label{tab:irreps}}
\end{table}

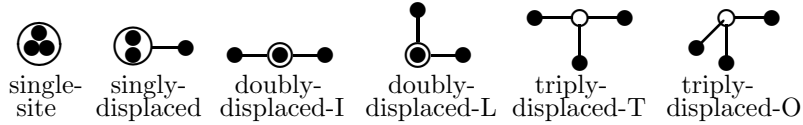
\begin{figure}[t]
\centerline{ \hspace{-4mm}
\raisebox{0mm}{\setlength{\unitlength}{1mm}
\thicklines
\begin{picture}(16,10)
\put(8,6.5){\circle{6}}
\put(7,6){\circle*{2}}
\put(9,6){\circle*{2}}
\put(8,8){\circle*{2}}
\put(4,0){single-}
\put(5,-3){site}
\end{picture}} \hspace{-6mm}
\raisebox{0mm}{\setlength{\unitlength}{1mm}
\thicklines
\begin{picture}(16,10)
\put(7,6.2){\circle{5}}
\put(7,5){\circle*{2}}
\put(7,7.3){\circle*{2}}
\put(14,6){\circle*{2}}
\put(9.5,6){\line(1,0){4}}
\put(4,0){singly-}
\put(2,-3){displaced}
\end{picture}}  \hspace{-5mm}
\raisebox{0mm}{\setlength{\unitlength}{1mm}
\thicklines
\begin{picture}(20,8)
\put(12,5){\circle{3}}
\put(12,5){\circle*{2}}
\put(6,5){\circle*{2}}
\put(18,5){\circle*{2}}
\put(6,5){\line(1,0){4.2}}
\put(18,5){\line(-1,0){4.2}}
\put(6,0){doubly-}
\put(4,-3){displaced-I}
\end{picture}}  
\raisebox{0mm}{\setlength{\unitlength}{1mm}
\thicklines
\begin{picture}(20,13)
\put(8,5){\circle{3}}
\put(8,5){\circle*{2}}
\put(8,11){\circle*{2}}
\put(14,5){\circle*{2}}
\put(14,5){\line(-1,0){4.2}}
\put(8,11){\line(0,-1){4.2}}
\put(4,0){doubly-}
\put(1,-3){displaced-L}
\end{picture}}   \hspace{-4mm}
\raisebox{0mm}{\setlength{\unitlength}{1mm}
\thicklines
\begin{picture}(20,12)
\put(10,10){\circle{2}}
\put(4,10){\circle*{2}}
\put(16,10){\circle*{2}}
\put(10,4){\circle*{2}}
\put(4,10){\line(1,0){5}}
\put(16,10){\line(-1,0){5}}
\put(10,4){\line(0,1){5}}
\put(4,0){triply-}
\put(1,-3){displaced-T}
\end{picture}} \hspace{-4mm}
\raisebox{0mm}{\setlength{\unitlength}{1mm}
\thicklines
\begin{picture}(20,12)
\put(10,10){\circle{2}}
\put(6,6){\circle*{2}}
\put(16,10){\circle*{2}}
\put(10,4){\circle*{2}}
\put(6,6){\line(1,1){3.6}}
\put(16,10){\line(-1,0){5}}
\put(10,4){\line(0,1){5}}
\put(4,0){triply-}
\put(2,-3){displaced-O}
\end{picture}}  }
\vspace*{8pt}
\caption{The spatial arrangments of the extended three-quark baryon
operators used. Smeared quark fields are
shown by solid circles, line segments indicate
gauge-covariant displacements, and each hollow circle indicates the location
of a Levi-Civita color coupling.  For simplicity, all displacements
have the same length in an operator.
\label{fig:two}}
\end{figure}

Baryons are expected to be rather large objects, so local operators
will not suffice.  Our approach to constructing spatially-extended operators 
is to use covariant displacements of the quark fields along the 
links of the lattice.
Displacements in different directions are used to build up the appropriate
orbital structure, and displacements of different lengths can build up
the needed radial structure.  All our three-quark baryon operators are
superpositions of gauge-invariant, translationally-invariant terms of the form
\begin{equation}
\Phi^{ABC}_{\alpha\beta\gamma,ijk}(t)
=  \sum_{\vec{\bm{x}}}  \varepsilon_{abc}\ (\tilde{D}^{(p)}_i\!\tilde{\psi}(\vec{\bm{x}},t))^A_{a\alpha} 
  \ (\tilde{D}^{(p)}_j\!\tilde{\psi}(\vec{\bm{x}},t))^B_{b\beta}
  \ (\tilde{D}^{(p)}_k\!\tilde{\psi}(\vec{\bm{x}},t))^C_{c\gamma},
\end{equation}
where $A,B,C$ indicate quark flavor, $a,b,c$ are color indices,
$\alpha,\beta,\gamma$ are Dirac spin indices,
$\tilde{\psi}$ indicates a smeared quark field, and
$\tilde{D}^{(p)}_j$ denotes the smeared $p$-link covariant displacement
operator in the $j$-th direction.  The smearing of the quark and
gauge field will be discussed later.   There are six different
spatial orientations that we use, shown in Fig~\ref{fig:two}.  The
singly-displaced operators are meant to mock up a diquark-quark coupling,
and the doubly-displaced and triply-displaced operators are chosen since
they favor the $\Delta$-flux and $Y$-flux configurations, respectively.

Next, the $\Phi^{ABC}_{\alpha\beta\gamma, ijk}$ are combined into 
{\em elemental} operators $B^F_a(t)$ having the appropriate flavor
structure characterized by isospin, strangeness, {\it etc.}   We work in 
the $m_u=m_d$ (equal $u$ and $d$ quark masses) approximation, and thus,
require that the elemental operators have definite isospin, that is, they 
satisfy appropriate commutation relations with the isospin operators
$\tau_3,\tau_+,\tau_-$.  Since we plan to compute full correlation matrices,
we need not be concerned with forming operators according to an $SU(3)$ 
flavor symmetry.  
Maple code which manipulates Grassmann fields was used to identify
maximal sets of linearly independent elemental operators.  The operators
used are shown in Table~\ref{tab:elemental}, and the numbers of such
independent operators are listed in this table as well.

\begin{table}[t]
\tbl{Elemental operators for various baryons (left).  The numbers of 
 linearly independent elemental operators of each spatial kind (right).}
{\begin{tabular}{cc|crrrr}\hline
{} &{} &{} &{} &{} &{} &{}\\[-1.5ex]
Baryon & Operator & Spatial Type & $\Delta,\Omega$ & $N$ & $\Sigma,\Xi$ 
   & $\Lambda$\\[1ex] \hline  {} &{} &{} &{} &{} &{} &{}\\[-1.5ex]
$\Delta^{++}$ & $\Phi^{uuu}_{\alpha\beta\gamma,ijk}$ & single-site        &  20  & 20  &  40   & 24  \\[1ex]
$\Sigma^{+}$ & $\Phi^{uus}_{\alpha\beta\gamma,ijk}$ & singly-displaced   &  240 & 384 &  624  & 528 \\[1ex]
$N^{+}$ & $\Phi^{uud}_{\alpha\beta\gamma,ijk}-\Phi^{duu}_{\alpha\beta\gamma,ijk}$ &\hspace{-3mm} doubly-displaced-I &  192 & 384 &  576  & 576 \\[1ex]
$\Xi^{0}$ & $\Phi^{ssu}_{\alpha\beta\gamma,ijk}$ &\hspace{-3mm} doubly-displaced-L &  768 & 1536&  2304 & 2304\\[1ex]
$\Lambda^{0}$ & $\Phi^{uds}_{\alpha\beta\gamma,ijk}-\Phi^{dus}_{\alpha\beta\gamma,ijk}$ & triply-displaced-T &  768 & 1536&  2304 & 2304\\[1ex]
$\Omega^{-}$ & $\Phi^{sss}_{\alpha\beta\gamma,ijk}$ & triply-displaced-O &  512 & 1024&  1536 & 1536\\[1ex] \hline
\end{tabular}
\label{tab:elemental}}
\end{table}

The final step in our operator construction is to apply group-theoretical 
projections to obtain operators which transform irreducibly under all lattice
rotation and reflection symmetries:
\begin{equation}
  B_a^{\Lambda\lambda F}\!(t)\!
 = \frac{d_\Lambda}{g_{O_h}} \sum_{R\in O_h}
  \Gamma^{(\Lambda)}_{\lambda\lambda}(R)
   \ U_R\ B^F_a(t)\ U_R^\dagger,
\label{eq:project}\end{equation}
where $\Lambda$ refers to an $O_h$ irrep, $\lambda$ is the irrep row, $g_{O_h}$ is
the number of elements in $O_h$, $d_\Lambda$ is the dimension of the $\Lambda$ irrep,
$\Gamma^{(\Lambda)}_{mn}(R)$ is a $\Lambda$ representation matrix corresponding to 
group element $R$, and $U_R$ is the quantum operator which implements the symmetry
operations.  The projections in Eq.~(\ref{eq:project}) are carried out
using computer software written in the Maple\cite{maple} symbolic 
manipulation language.

\section{Field smearing and operator pruning}
\label{sec:smear}

For single-site (local) hadron operators, it is well known that the use
of spatially-smeared quark fields is crucial.  For extended baryon operators,
one expects quark-field smearing to be equally important, but the relevance
and interplay of link-field smearing is less well known. Thus, we decided that
a systematic study of both quark-field and link-variable smearing was warranted.

The link variables were smeared $U\rightarrow\tilde{U}$ using the analytic stout 
link method of Ref.~\refcite{stout}.  There are two tunable parameters, the number 
of iterations $n_\rho$ and the staple weight $\rho$.  For the quark-field, we 
employed gauge-covariant Gaussian smearing
\begin{eqnarray}
 \tilde{\Psi}(x)&=&\left(1+\frac{\sigma_s^2}{4n_\sigma}
 \tilde{\Delta}\right)^{n_\sigma}\Psi(x),\\
\tilde\Delta\Psi(x)&=&\sum_{k=\pm1,\pm2,\pm3}\left(\tilde{U}_k(x)\Psi(x+\hat{k})-\Psi(x)\right),
\end{eqnarray}
where $\tilde\Delta$ denotes the smeared three-dimensional gauge-covariant Laplacian.
The two parameters to tune in this smearing procedure are the smearing radius $\sigma_s$ and 
the integer number of iterations $n_{\sigma}$.

\begin{figure}[t]
\includegraphics[width=4.25in, bb=0 0 567 559]{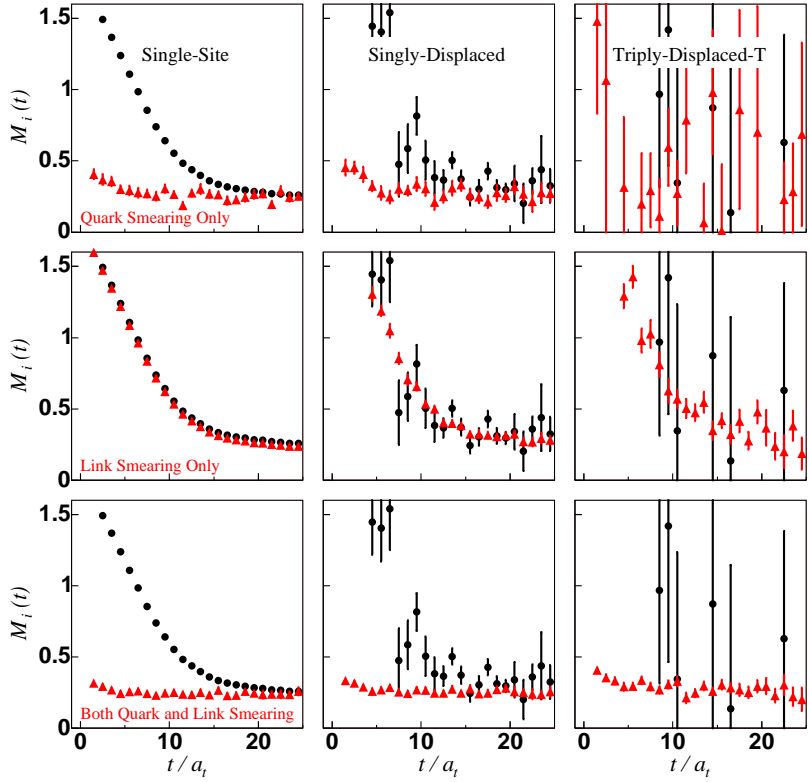}
\caption{Effective masses $M(t)$ for unsmeared (circles) and smeared 
(triangles) operators $O_{SS},\ O_{SD},\ O_{TDT}$. 
Top row: only quark-field smearing $n_\sigma=32,\ \sigma_s=4.0$ is used. Middle row: 
only link-variable smearing $n_\rho=16,\ n_\rho\rho=2.5$ is applied.  
Bottom row: both quark and link smearing $n_\sigma=32,\ \sigma_s=4.0, 
\ n_\rho=16,\ n_\rho\rho=2.5$ are used, dramatically improving the signal for all
three operators. Results are based on 50 quenched configurations on a 
$12^3\times 48$ anisotropic lattice using the Wilson action with $a_s \sim 0.1$ fm,
$a_s/a_t \sim 3.0$. The quark mass was chosen so that $m_{\pi} \simeq 700~{\rm MeV}$.
\label{fig:meff-smear}}
\end{figure}

\begin{figure}[t]
\includegraphics[width=4.25in, bb=17 26 591 584]{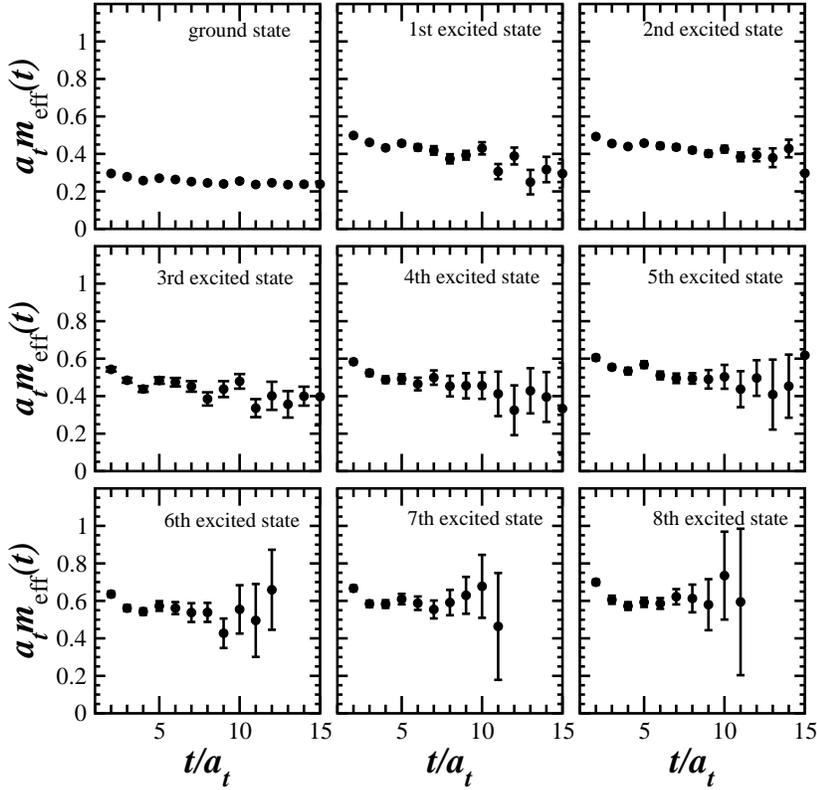}
\caption{Principal effective masses for the lowest-lying \textit{nine} states
in the $G_{1g}$ nucleon channel.  Results are based on 100 quenched configurations 
on a $12^3\times 48$ anisotropic lattice using the Wilson action with 
$a_s \sim 0.1$ fm, $a_s/a_t \sim 3.0$. The quark mass was chosen so 
that $m_{\pi} \simeq 700~{\rm MeV}$.  Quark-field smearing 
$n_\sigma=32,\ \sigma_s=3.0$ and link-smearing $n_\rho=16,\ n_\rho\rho=2.5$
are used.  These principal effective masses come from a $10\times 10$ correlation
matrix, where 2 singly-displaced, 5 doubly-displaced-I, 2 doubly-displaced-L,
and 1 triply-displaced-T operators were chosen.
\label{fig:principaleffmasses}}
\end{figure}

For our tests of the efficacy of quark-field and gauge-link smearing, correlators
were computed for three particular nucleon operators: a single-site operator $O_{SS}$ in 
the $G_{1g}$ irreducible representation of the cubic point group, a singly-displaced
operator $O_{SD}$ with a particular choice of each Dirac index, and a triply-displaced-T
operator $O_{TDT}$ with a specific choice of each Dirac index.  Our findings are summarized
in Fig.~\ref{fig:meff-smear}.  The top row shows that applying only quark-field smearing
to the three selected nucleon operators significantly reduces couplings
to higher-lying states, but the displaced operators remain excessively
noisy.  The second row illustrates that including only link-field smearing
substantially reduces the noise, but does not appreciably alter the effective
masses themselves.  The bottom row shows dramatic improvement from reduced
couplings to excited states and dramatically reduced noise when both
quark-field and link-field smearing is applied, especially for the extended
operators.   The ``best'' quark-field smearing parameters $n_{\sigma}$ and $\sigma$ 
were determined by requiring that the effective mass for the three chosen operators
reach a plateau as  close to the source as possible.  The gauge-link smearing parameters
were tuned so as to minimize the noise in the effective masses.  One interesting point
we also learned was that the preferred link-smearing parameters determined from the static 
quark-antiquark potential produced the smallest error in the extended baryon operators.

The computation of correlation matrices using hundreds of operators is not
feasible, so it is necessary to ``prune'' out unnecessary operators.
The first step in this pruning is to examine the effective masses of the
diagonal elements of the correlation matrices to identify and eliminate
noisy operators.  Keeping only operators with small statistical uncertainties
yields a set of about forty to fifty operators in each symmetry channel.  We
then computed the correlation matrix of this reduced set of operators, examining
whether further reductions to the operator sets could be made without increased
contamination in the principal effective masses.  These computations
are still ongoing, but preliminary results are shown in 
Fig.~\ref{fig:principaleffmasses}.  This figure shows that it is possible
to extract at least \textit{nine} levels in a given symmetry channel, a feat
which has never before been accomplished.  Demonstrating that this number
of energy levels can be reliably extracted is an important milestone in our
long-term project.

\section{Conclusion}
\label{sec:conclude}

We have outlined a program to study the resonance spectrum in lattice
QCD.  The use of the variational method and the need to isolate
several energy levels in each channel require a sufficiently broad
basis of operators.  Having developed suitable group-theory methods to
project operators onto the irreducible representations of the cubic
group, and having examined the efficacy of both quark- and
gauge-link-smearing, we are now identifying a more limited set of
operators that we will employ in a large-scale study of the hadron
spectrum.  Our methods are applicable not only to baryons, but also to
mesons, tetra-quark and pentaquark systems, and to states with excited
glue.  Only by performing such a program can we hope to
identify the states of QCD, and in particular their spins and
parities, in the continuum limit.  Ultimately, when quark loops are 
included at realistically light quark masses, multi-hadron (baryon-meson) 
operators must be included in our correlation matrices, and finite-volume
techniques will need to be employed to ferret out the baryon resonances
from uninteresting scattering states.  We are currently exploring
different ways of building such operators. 

This work was supported by the U.S.~National Science Foundation
through grants PHY-0354982, PHY-0510020, and PHY-0300065, and by the 
U.S.~Department of Energy under contracts DE-AC05-84ER40150 and 
DE-FG02-93ER-40762. Computations were performed using the 
\textit{Chroma} software package\cite{chroma}.

\end{document}